\documentstyle{l-aa}
\input epsf

\begin{document}

\thesaurus{08    	
             	(08.14.1; 	
		08.02.3; 	
  		08.16.6; 	
		08.16.7 SAX J1808.4-3658; 
	       	13.25.1 	
	       	13.25.3 	
		)}

\title{The millisecond X--ray pulsar/burster SAX J1808.4-3658: the
outburst light curve and  the power law spectrum.}

\author{M.Gilfanov\inst{1,2}, M.Revnivtsev\inst{2,3}, 
R.Sunyaev\inst{1,2} and E.Churazov\inst{1,2}}

\offprints{M. Gilfanov}

\institute{Max-Planck-Institut f\"ur Astrophysik,
Karl-Schwarzschild-Str. 1, 85740 Garching bei Munchen, Germany 
\and
Space Research Institute,Russian Academy of Sciences, Profsoyuznaya
84/32, 117810 Moscow, Russia 
\and
visiting Max-Planck-Institut f\"ur Astrophysik
}

\date{\today}

\maketitle

\begin{abstract}
The X--ray light curve and broad band spectral properties of the
millisecond X--ray pulsar/burster SAX J1808.4-3658
are reported. In the course of RXTE observations in April--May 1998 the
3--150 keV luminosity of the source decreased  by a factor of $\sim 100$
from the peak value of  $\sim 9\times 10^{36}$ erg/s (for a 4 kpc distance).
However, the spectrum was remarkably stable and maintained a roughly power
law shape with a photon index of $\sim 2$ without
a strong high energy cut-off below 100 keV, similar to that sometimes
observed in the low spectral state of X--ray bursters. 
An approximation of the averaged spectrum with an exponentially
cut-off power  law with a superimposed reflected component yields the
90\% lower limit on the e-folding energy $E_{f}\ga 270$ keV. 
We speculate that Comptonization on the bulk
motion in the radiation dominated shock might be a possible mechanism
of spectral formation.
The decaying part of the X--ray light curve was cut off abruptly
at  luminosity  $\sim {\rm few}\times 10^{35}$ erg/s.
Such behavior might be due to centrifugal inhibition of accretion
(transition to the ``propeller'' regime) as well as to disk instability.
In either case an upper limit on the neutron star
magnetic field strength is $B\la{\rm few}\times 10^7$ Gauss.


\end{abstract}

\markboth{M.Gilfanov et al.: The millisecond X--ray pulsar/burster SAX
J1808.4-3658.}{}

\section{Introduction}

The transient X-ray source SAX J1808.4--3658 was discovered in September
1996 using the Wide Field Camera (WFC) aboard BeppoSAX 
(\cite{saxpaper}).  The source was not detected by WFC in August and October
1996 or by  MIR--KVANT/TTM (N.Alexandrovich \& K.Borozdin, private
communication)  7 years before in March and August--September 1989 with the
upper limits  of $\sim   1/50$ and $\sim 1/10$ of the September 1996
peak flux.  Two Type I X--ray bursts were detected by WFC in
September 1996. The analysis of the X--ray bursts gave 4 kpc source
distance for which the 0.4--10 keV  persistent luminosity 
was $6\cdot10^{36}$ erg/s. According to ASM/RXTE data the
1996 outburst lasted for $\sim20$ days.

Recently the Rossi X--Ray Timing Explorer detected a new outburst 
from SAX J1808.4--3658 (Marshall et al. 1998). 
Coherent pulsations with a period of 2.49 msec (Wijnands \& van der Klis
1998) and Doppler shift due to orbital motion with a period of $2^h$
(Chakrabarty \& Morgan 1998) were discovered. The analysis of the RXTE/HEXTE
data from  April 
11 and 13  revealed a remarkable power law spectrum of the
source in the 15--120 keV band with a photon index of $\approx 2$
(\cite{hexte}). The optical counterpart of SAX J1808.4--3658 
brightened by $>3.4^m$ and reached $m_V=16.6$ on April 18
(\cite{optical1}). \cite{optical2} 
detected a roughly sinusoidal modulation of $0.12^m$  in
the V--band with a $2^h$ binary system period.

\begin{table*}
\small
\caption{SAX J1808.4--3658. Parameters of spectral approximation of
PCA and HEXTE data by a power law model
\label{fit}} 
\tabcolsep=0.19cm
\begin{tabular}{lcrrccccccc}
\hline
Date& Time, UT&\multicolumn{2}{c}{Expos.$^1$,sec}&\multicolumn{2}{c}{PCA}&\multicolumn{2}{c}{HEXTE}&\multicolumn{3}{c}{PCA+HEXTE}\\
1998& &\multicolumn{3}{l}{~PCA~HEXTE$^2$~~~~~~~$\alpha$}&$F_{3-25keV}^3$&$\alpha$&$F_{20-150 keV}^{3,4}$&$\alpha$&$F_{3-150keV}^{3,5}$&$\chi^2_r$\\
\hline
11/04 & 19:50--21:44 &  2974& 2041 & $2.00\pm0.01$ & $245.4\pm0.8$ &$2.13\pm0.09$ &$138.8\pm9.6$ &$2.01\pm0.01$ &$450.0\pm3.1$&1.50\\
13/04 & 01:52--02:23 &  1375&  949 & $1.97\pm0.01$ & $224.2\pm0.8$ &$2.13\pm0.12$ &$153.9\pm4.2$ &$1.97\pm0.01$ &$423.4\pm3.3$&1.36\\
16/04 & 17:19--23:01 &  9266& 8620 & $1.95\pm0.01$ & $145.3\pm0.5$ &$2.03\pm0.06$ &$ 99.9\pm4.8$ &$1.96\pm0.01$ &$279.3\pm1.8$&1.43\\
17/04 & 01:42--04:39 &  5675& 4018 & $1.95\pm0.01$ & $140.8\pm0.5$ &$2.10\pm0.08$ &$ 91.5\pm5.7$ &$1.96\pm0.01$ &$270.4\pm1.9$&1.41\\
18/04 & 03:09--09:19 & 13980& 9179 & $1.96\pm0.01$ & $123.6\pm0.4$ &$2.06\pm0.05$ &$ 82.9\pm3.7$ &$1.96\pm0.01$ &$235.9\pm1.6$&1.48\\
18/04 & 12:34--01:03 & 25285&17056 & $1.96\pm0.01$ & $118.5\pm0.4$ &$2.08\pm0.05$ &$ 77.5\pm3.1$ &$1.97\pm0.01$ &$225.5\pm1.5$&1.54\\
20/04 & 21:03--23:09 &  5120& 3557 & $1.98\pm0.01$ & $ 98.1\pm0.3$ &$2.11\pm0.13$ &$ 59.5\pm6.4$ &$1.99\pm0.01$ &$182.7\pm1.5$&1.49\\
23/04 & 15:53--23:21 & 16493&11135 & $2.01\pm0.01$ & $ 81.0\pm0.3$ &$2.03\pm0.09$ &$ 55.7\pm4.0$ &$2.01\pm0.01$ &$147.8\pm1.1$&1.59\\
24/04 & 16:08--23:22 &  8547&10426 & $2.03\pm0.01$ & $ 77.6\pm0.3$ &$2.12\pm0.10$ &$ 47.1\pm3.8$ &$2.04\pm0.01$ &$138.5\pm1.1$&1.49\\
25/04 & 14:23--21:41 & 15045&10120 & $2.04\pm0.01$ & $ 73.1\pm0.3$ &$2.08\pm0.11$ &$ 45.7\pm4.0$ &$2.05\pm0.01$ &$129.5\pm1.0$&1.60\\
26/04 & 16:01--23:28 & 15819&10748 & $2.05\pm0.01$ & $ 48.5\pm0.2$ &$2.11\pm0.16$ &$ 27.6\pm3.6$ &$2.05\pm0.01$ &$ 85.6\pm0.8$&1.36\\
27/04 & 14:30--19:31 &  8765& 6006 & $2.11\pm0.02$ & $ 24.8\pm0.2$ &$2.15\pm0.44$ &$ 15.8\pm5.1$ &$2.11\pm0.01$ &$ 41.8\pm0.6$&1.22\\
29/04 & 14:24--18:58 &  8536& 5735 & $2.10\pm0.02$ & $ 19.9\pm0.1$ &$2.39\pm0.61$ &$ 12.7\pm5.2$ &$2.10\pm0.02$ &$ 33.8\pm0.6$&1.04\\
02/05$^6$ & 01:26--02:59 &3259&1988 & $1.86\pm0.13$ & $  1.6\pm0.1$ &$ 2.0^7  $ &$ \la4.7^8$ &$1.86\pm0.13$ &$  3.4\pm0.6$&0.70\\
03/05$^6$ & 19:10--21:01 &4072&2790 & $2.29\pm0.08$ & $  3.3\pm0.1$ &$ 2.0^7       $ &$ \la4.6^8$ &$2.29\pm0.07$ &$  4.9\pm0.4$&0.70\\
\hline
\end{tabular}
The errors are 90\% confidence;
$^1$ -- dead time corrected;
$^2$ -- on source, sum of two clusters of detectors;
$^3$-- the energy flux in $10^{-11}$ erg/sec/cm$^2$;
$^4$-- original HEXTE normalization
$^5$-- the HEXTE normalization was adjusted to match the PCA data;
$^6$-- low source count rate ($\sim 5-10\%$ of the PCA
background, see text); 
$^7$-- fixed;
$^8$-- 90\% upper limit;
\end{table*}

\section{Observations and data reduction.}

We used the target--of--opportunity public domain data 
of the PCA and HEXTE instruments aboard RXTE (\cite{rxte}).  
The data were analyzed according to RXTE Cook Book recipes using FTOOLS,
version 4.1, tasks. The VLE and L7 background models were used for 
the PCA data acquired before and after May 2 respectively. 
Due to the low source count rate ($\sim 5-10\%$ of the PCA background)
the PCA spectra and the best fit parameters for the May 2 and 3
observations 
might be affected by the background subtraction errors. The May 6 data
($\sim 3-5\%$ of the background) were excluded from the
spectral analysis. A systematic error of 1\% of the source count rate and
2\% of the PCA background count rate was added quadratically to the
statistical error. The ASM light curve was retrieved from
http://space.mit.edu/XTE/ASM\_lc.html.

\section {Results.}

\begin{figure}
\epsfxsize=8.7cm
\epsffile{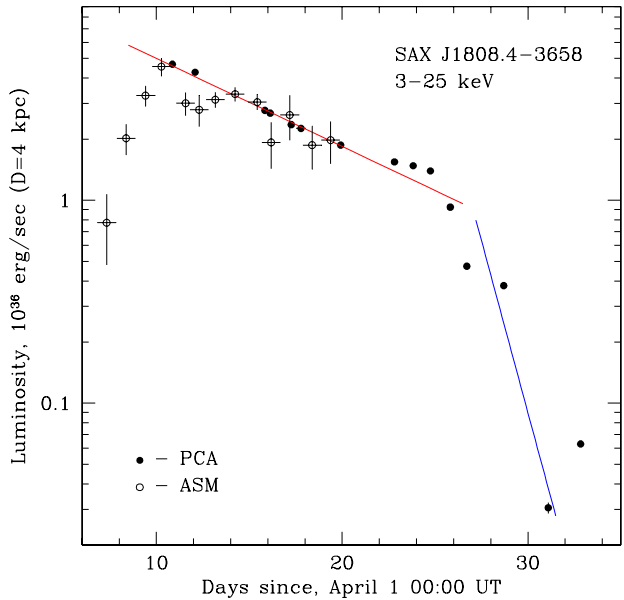} 
\caption{The 3--25 keV light curve of SAX J1808.4--3658. The PCA fluxes are
those from Table \ref{fit}, the ASM count rate was converted to 3--25 keV
energy flux assuming a Crab like spectrum. The solid lines 
are $L_X\propto e^{-t/10^d}$ and $L_X\propto e^{-t/1.3^d}$.
\label{lc}}   
\end{figure}

The 3--25 keV light curve and the spectra  of SAX J1808.4--3658
are shown in Figs.\ref{lc} and \ref{spezoo}. In order to characterize
the broad band spectral properties at different luminosity the power
law model was used. The best fit parameters are listed in Table
\ref{fit}. 
The pulsation with $\approx 2.5$ msec period was
detected in all observations between April 11 -- 29 and on May 3 with
the relative rms of $\approx 4-7\%$ increasing slightly as luminosity 
decreased.

A power law fit to the 3--150 keV spectrum averaged over 
April 11--25 gives a value of reduced $\chi^2_r=2.0$ for 297
dof. The deviations are mostly due to PCA data: $\chi^2_r=4.4$ (50 dof) and
$\chi^2_r=1.3$ (246 dof) for PCA and HEXTE respectively.  
\cite{hs} noted that the addition of an emission line at $\sim 6.8$
keV with an equivalent width $\approx 50$ eV, a black body component at
lower energy and a high energy cut off above $\approx 30$ keV with
an e--folding energy of $E_f\approx 127$ keV significantly improved the
quality of the fit in terms of the $\chi^2$ statistics. 

The significant brightening of the source in the optical band
(\cite{optical1}) plus  an iron K$-\alpha$ line and a broad hump near
$\sim 30$ keV  in X--ray band (Fig.\ref{spezoo}) indicate 
the presence of a reflected component  which might result in some
apparent steepening of the  spectrum above 30 keV. An accurate account
for the effect of reflection  requires knowledge of the geometry and
ionization state of the reflecting material and accurate
cross-calibration of the PCA and HEXTE instruments. 
In order to estimate the shape of the intrinsic source spectrum
we fit the April 11-25 averaged 3--150 keV data with an exponentially
cut-off power law reflected from neutral material (pexrav model in XSPEC).
The equivalent width of the line, $\sim 50$ eV,
suggests that the solid angle subtended by the reflecting media is
$\Omega/2\pi\sim 0.4$ (\cite{bstrefl,fabian}), assuming a solar
abundance of iron and a fiducial value of the disk inclination angle
of $\theta=60\degr$. 
The  reflection scaling factor, fixed at this value,
the best fit parameters for the intrinsic spectrum are: photon index
$\alpha\approx 2.00\pm0.02$ and e-folding energy $E_{f}\approx 
365_{-95}^{+175}$ keV (90\%  errors) with
$\chi^2_r=1.11$ (292 dof). The hydrogen column density was fixed at the
Galactic value $NH=1.3\cdot 10^{21}$ cm$^{-2}$. A variation of the the
reflection scaling factor and the inclination angle does not 
qualitatively change $\alpha$ and $E_f$:
$\alpha\approx 1.96\pm0.02$, $E_{f}\approx 255_{-50}^{+80}$ keV for 
$\Omega/2\pi\sim 0.3$, $\theta=60\degr$ and 
$\alpha\approx 2.03\pm0.02$, $E_{f}\approx 395_{-100}^{+90}$ keV for 
$\Omega/2\pi\sim 0.3$, $\theta=30\degr$.

The results of the RXTE observations of SAX J1808.4--3658 can be
summarized as follows: 
\begin{enumerate}
\item The 3--100 keV spectrum maintained an approximate power law shape
$I_\nu\propto\nu^{-2}$ as luminosity decreased by a factor of $\sim 100$
(Table \ref{fit}, Fig.\ref{spezoo}).  
Under the abovementioned  assumptions about the reflected component, the
lower limit on the exponential cut-off in the intrinsic spectrum
averaged over April 11--25 is $E_{f}\ga 270$ keV. 

The spectrum differs substantially from the low state 
spectra of the X--ray bursters often  observed at 
$L_X\sim 10^{37}$ erg/s (3--150 keV; e.g. GX354--0, 4U1705--44,
Fig.\ref{spezoo}). It is similar to the spectra of some
X--ray bursters  at sufficiently low luminosity $L_X\la
(1-2)\cdot 10^{36}$ erg/s (e.g. 4U1608--522 and Aql X--1,
Fig.\ref{spezoo}).  
\item The decaying part of the X--ray light curve has a sharp cut off
below $\sim 10^{36}$ erg/s (4 kpc distance is assumed throughout the paper
unless mentioned otherwise) -- the luminosity has dropped by 
a factor of $\sim20$ within $\la 3$ days. The rise time was also short: 
$\la 3$ days.
\item The total energy released in the 3--150 keV band was $\sim
8\times 10^{42}$ erg, corresponding to an accreted mass of
$\sim 2\times 10^{-11} M_{\sun}$. The peak value of the mass
accretion rate was $\dot{M}\sim 7\times 10^{-10} M_{\sun}/yr$.
These estimates however do not take into account the energy radiated
below 3 keV and above 150 keV which could be non negligible.
\item The $I_\nu\propto\nu^{-2}$ power law spectrum does not continue
to the IR band -- $\frac{L_{IR}}{L_X}\sim 10^{-2}$ (IR data -- from
\cite{optical1}).  
\end{enumerate}

\section{Discussion.}

\begin{figure*}
\hbox{
\epsfxsize=8.7cm
\epsffile{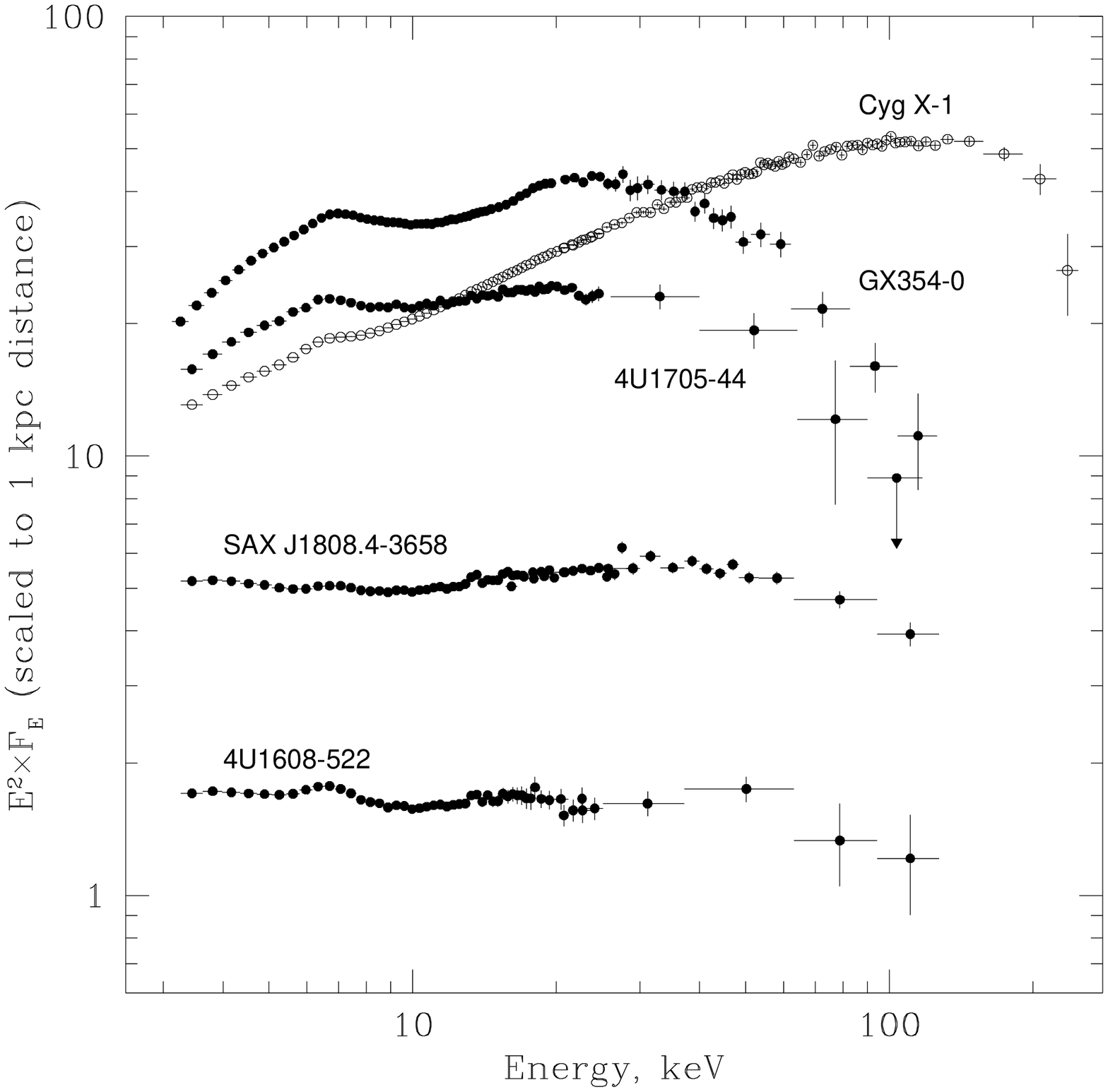} 
\hspace{2mm}
\epsfxsize=8.7cm
\epsffile{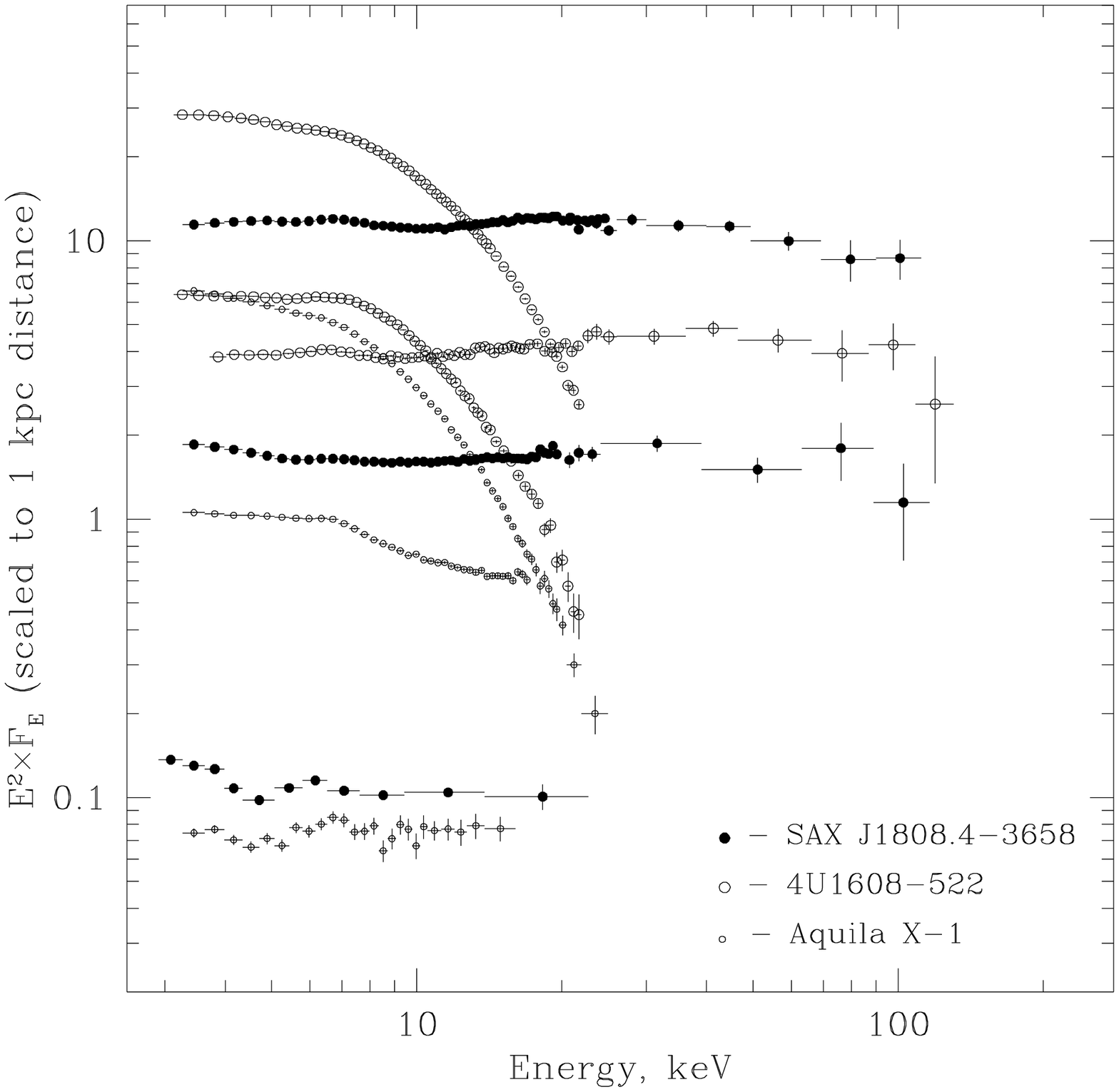} 
}
\caption{{\em Left:} The  averaged (April 11--25) spectrum of SAX
J1808.4--3658 and the spectra of several X--ray bursters and
black hole Cyg X--1 in the low spectral state. {\em Right:} The
spectra of SAX J1808.4--3658 obtained on April 11--13, 26--29 and May
2-3 in comparison with the spectra of other X--ray bursters at
different luminosity: 4U1608--522 (Spring 1996 outburst) and Aquila
X--1 (Spring 1997). All spectra were obtained using RXTE public data.
\label{spezoo}}   
\end{figure*}

Theories of the spectral formation in accreting pulsars
with a strong magnetic field predict a change of the physical
conditions near the surface of the neutron star at 
$L_X\sim{\rm few}\times 10^{36}$ erg/sec. 
In the case of X--ray bursters strong spectral changes are associated  
with the transition from the soft to the hard spectral state at 
$L_X\sim 10^{36}-10^{37}$ erg/sec (Fig.\ref{spezoo}). 
We therefore expected to observe spectral evolution as luminosity decreased  
by a factor of  $\sim 100$. Remarkably no significant spectral changes were
detected (Table \ref{fit}, Fig.\ref{spezoo}). 
This, together with the detection of X--ray pulsations of roughly
the same relative amplitude, implies that essentially the same
mechanism is responsible for the spectral formation in a broad
luminosity range.

\subsection{A rotation powered pulsar ?}

Stability and shape of the spectrum of SAX
J1808.4--3658  suggests considering a possibility,
that the 1998 outburst was powered by the standard pulsar
emission mechanism. For  magnetic dipole emission the field
strength required to generate X--ray flux $F_X$ [erg/s/cm$^2$]
is  
$ 
B\sim 2.2\cdot 10^9
D_{\rm kpc}
R_6^{-3}
P_{2.5}^4
\left(\frac{\beta}{0.1}\right)^{-1/2}
\left(\frac{F_X}{4\cdot 10^{-9}}\right)^{1/2}
$
Gauss, 
where $D_{\rm kpc}$ is the distance in kpc, $R_6$ --
the neutron star radius in $10^6$ cm, $P_{2.5}$ -- rotation period divided
by 2.5 msec, 
$\beta$ -- the efficiency of the production of X--ray
photons. 
On the other hand  the 1996 outburst was accretion powered 
as the Type I X--ray bursts were detected. 
For the accretion to proceed uninhibited by the centrifugal force due to 
the rotating magnetosphere of the neutron star the  accretion rate
would be 
$
\dot{M}\ga 5\cdot 10^{17} B^2_9 R^{16/3}_6 
M_{1.4}^{-5/3} P_{2.5}^{-7/3}$ 
g/s i.e. 
$\dot{M}\ga 2.5\cdot 10^{18} D_{\rm kpc}^2$
for $B\sim 2.2\cdot 10^9~D_{\rm kpc}$ Gauss.
Therefore, independent of the source distance, the 2--10 keV
flux observed in 1996 ($2\cdot 10^{-9}$
erg/s/cm$^2$) would correspond to  $\sim 10^{-4}$ of $\dot{M}c^2$. 
Such a low efficiency of production of X--ray
photons is unlikely for an accreting neutron
star. Moreover, such a high $\dot{M}$  would lead to a steady
nuclear burning of the accreting matter, which
contradicts the detection of the Type I X--ray bursts.
Thus,  the standard pulsar emission mechanism can be
ruled out.

\subsection{The X--ray light curve.}

The X--ray light curve is  similar to the
outburst profiles of two very different types of accreting sources --
the neutron star soft X--ray transients (e.g. \cite{aql,aqlsax}) 
and Type B outbursts in dwarf Novae (\cite{cvrev}). An abrupt
cut-off of the light curve is also common for the black hole
X--ray transients (\cite{bhrev}). 
Two different interpretations of the abrupt decline of luminosity at the end
of the outburst are possible in the case of SAX J1808.4--3658:
\begin{enumerate} 
\item Closure of the centrifugal barrier at low
accretion rate when the magnetosphere reaches the corotation
radius  
$R_{co}\approx 31 M_{1.4}^{1/3} P_{2.5}^{2/3}$ km
and the source enters the ``propeller'' regime
(\cite{propeller}).  The cut-off luminosity provides an estimate of
the magnetic field as:  
$B\sim 3\cdot 10^7
M_{1.4}^{1/3} R_{6}^{-8/3} P_{2.5}^{7/6}
\left(\frac{L_X}{10^{35}}\right)^{1/2}
$ Gauss.
\item Disk instability. 
The decay time scale is defined by the propagation of
the cooling wave. A rough estimate for the SAX J1808.4--3658
binary system parameters is $\tau_d\sim
1.3\left(\frac{0.1}{\alpha}\right)$ days, $\alpha$ -- viscosity
parameter (\cite{cvrev}) which approximately accounts for the observed
decay time scale (Fig.\ref{lc}). The above expression for B
is an upper limit in this case. 

\end{enumerate}

\subsection{Spectral formation.}

A possible mechanism resulting in the formation of a power law 
spectrum with a slope $\sim 2$ 
might be Comptonization on the bulk motion in a radiation
dominated shock (Blandford \& Payne 1981, \cite{rdshock}). 
If, as indicated by the 
X--ray pulsation, there is non negligible disk--magnetosphere
interaction, the accreting matter is funneled onto the polar caps with
$\sim$free fall velocity. If the radiation energy flux $q$ exceeds the
critical value $q_c\sim$local Eddington flux, 
the radiation dominated shock is formed near the surface of the neutron star
at the polar regions (\cite{rdshock1,bs}), where $\ga 
2/3$ of the accretion energy is released. 
A naive estimate -- assuming that the magnetosphere is close
to corotation, the magnetic field inside the magnetosphere is not
distorted and the in-falling matter fills the magnetic funnel -- gives
the value of the critical luminosity 
$L_c\sim {\rm few}\times 10^{36} - 10^{37}$  
erg/sec, which exceeds by more than an order of magnitude the minimum
luminosity observed for  SAX J1808.4--3658.
However, an accurate calculation of $q/q_c$ requires detailed
knowledge of the disk--magnetosphere interaction and
the accretion flow pattern inside the magnetosphere.
It is important to note that in order for this mechanism to be
responsible for the spectral 
formation in SAX J1808.4--3658, self adjustment of the accretion funnel
to maintain $q/q_c\sim 1$ in a broad range of the accretion rates is
required.

Comptonization on the bulk motion in the radiation dominated shock
near the stellar surface in the equatorial area
might also occur in the case of the ``classical'' X--ray bursters
if the neutron star is smaller than the last marginally stable
orbit (\cite{smallns1,smallns}). 
The power law spectra with a slope $\sim
2$ up to $\sim 100$ keV are indeed observed for some of the
``classical'' X--ray bursters  
at low luminosity, $L_X\la 10^{36}$ erg/sec (Fig.\ref{spezoo}).

\section{Acknowledgments}
We thank F.Meyer, H.Ritter and H.Spruit for useful discussions and
the anonymous referee for helpful suggestions.
This research has made use of data obtained through the 
HEASARC Online Service, provided by the NASA/GSFC.

\end{document}